\documentclass[twocolumn,aps,showpacs]{revtex4-1}
\usepackage{graphicx}
\usepackage{times}
\usepackage{bm}
\begin{document}

\title{The anomalous $^{14}$C-dating $\beta$ decay problem revisited}

\author{Chong Qi}
\thanks{Email: chongq@kth.se}
\affiliation{Department of Physics, Royal Institute of Technology,
SE-10691 Stockholm, Sweden}

\begin{abstract}
The anomalous inhibition of  $^{14}$C-dating $\beta$ decay rate is restudied in terms of shell-model calculations in the $jj$ coupling scheme with both realistic and empirical Hamiltonians. It is seen that the accidental cancellation of the decay strength is dominated by the mixing effect of two configurations of the final state wave function, $|0p^{-2}_{1/2}\rangle$ and
$|0p_{3/2}^{-1}0p_{1/2}^{-1}\rangle$. By decomposing the effective interactions into different tensor components, it is clearly seen that the mixing is largely induced by the tensor force. The failure  of realistic calculations in reproducing
the inhibition may be related to its ill description of the monopole component rather than the tensor force.
\end{abstract}
\maketitle

The anomalously long $\beta$ decay half-life of $^{14}$C has been of special theoretical interest since the appearance of the nuclear shell model~\cite{Ing53,Talmi03}. The decay involves the $J^{\pi}=0^+$ ground state of $^{14}$C and the $J^{\pi}=1^+$ ground state of $^{14}$N and satisfies the selection rule for typical allowed Gamow-Teller (GT) transitions. However, the extracted transition amplitude from experimental half-life is thousands of times smaller than that of allowed transitions. The inhibition should be attributed to the accidental cancellation of certain components of the involved state wave functions that contribute to the transition. It was recognized that the tensor part of the nuclear force play an essential role in inducing the cancellation~\cite{Jan54,Rose68}. 

In the original paper of Jancovici and Talmi's~\cite{Jan54}, an unreasonably large tensor force was introduced to induce the cancellation. Later studies show that this problem can be rectified by redefining the radial dependence of the tensor component (for reviews, see Ref.~\cite{Talmi03}). One may expect that the shape and strength of the tensor force were confined in realistic nucleon-nucleon ({\it NN})  potentials which are determined by fitting {\it NN} scattering observables. But the studies of Zamick and collaborators~\cite{Zam66,Fay99,Zhen91} showed that the cancellation cannot be reproduced by calculations with realistic Hamiltonians~\cite{Kuo65} derived from microscopic {\it NN}  potentials like the Hamada-Johnston potential and the Bonn potential. This failure was also seen in recent calculations of Refs.~\cite{Holt08,Holt09,Holt10} with 
modern one-boson-exchange and chiral potentials. The problem may indicate that the tensor component of the in-medium {\it NN} interaction is much weaker than that of the bare potential~\cite{Zam66,Fay99,Zhen91}.

Recently Holt {\it et al.} claimed that the problem of realistic calculations in reproducing
the long  $\beta$ decay half-life of the $^{14}$C
can be solved by taking the Brown-Rho scaling in-medium modification~\cite{Holt08} or three-nucleon corrections~\cite{Holt09} of the interaction into account. However, these calculations changed dramatically the bulk properties of the effective interaction and did not shed any light on the role played by the tensor force. But perhaps even more serious is that the calculations fail completely in reproducing the cancellation by employing exactly the same approach but with a different  {\it NN} interaction~\cite{Holt10}.

The purpose of this work is to present a general calculation on the $^{14}$C $\beta$ decay property with available well-established empirical shell-model interactions and compare it with that of realistic interactions. By decomposing them into central, spin-orbit and tensor parts, we present a quantitative study on the role played by different components of  effective interactions in the cancellation of the $^{14}$C $\beta$ transition amplitude.  These may help in understanding the problem of realistic interactions in reproducing the cancellation and in illuminating the property the tensor force in the effective interaction and its possible in-medium modification effect. It may be also of interest to mention that the role played by the tensor force in the evolution of shell structure has also been extensively probed recently (see, e.g., Ref.~\cite{heyde}).

The wave functions of the $^{14}$C
($J^\pi=0^+$, $T=1$) and $^{14}$N ($J^\pi=1^+$, $T=0$) ground
states, denoted by $|\psi_i\rangle$ and $|\psi_f\rangle$,
respectively, can be well described as two holes occupying the $0p_{1/2}$ and $0p_{3/2}$ single-particle orbits assuming $^{16}$O as the inert core~\cite{Jan54,Rose68}. In the $jj$ coupling scheme,  the wave functions can be written as~\cite{Zam66}
\begin{eqnarray}
\nonumber|\psi_i\rangle &=& \kappa|0p^{-2}_{1/2}\rangle +
\eta|0p_{3/2}^{-2}\rangle,\\
|\psi_f\rangle &=& a|0p^{-2}_{1/2}\rangle
+b|0p_{3/2}^{-1}0p_{1/2}^{-1}\rangle+c|0p_{3/2}^{-2}\rangle,
\end{eqnarray}
where the $\kappa$ and $\eta$ and $a$, $b$ and $c$ denote the corresponding wave function amplitudes.
The GT transition matrix element is determined by
\begin{equation}
M({\rm GT})=\langle\psi_f||\bm{\sigma\tau}||\psi_i\rangle =
\sqrt{\frac{2}{3}}\left[\kappa(a+2b)+\eta(\sqrt{2}b-\sqrt{5}c)\right].
\end{equation}

In evaluating above wave function amplitudes and the transition amplitude $M({\rm GT})$, one may start from empirical as well as realistic shell-model interactions.
It has been established that realistic interactions derived from
bare {\it NN} potentials are in general close to empirical ones
which are obtained by fitting experimental data~\cite{Jensen,Zuker,Bro06,Honma}. This can also be seen from Table~\ref{table1} where the diagonal matrix elements of some well-defined empirical $p$-shell interactions~\cite{ck,van88,wbt} and those of the realistic interaction of Ref.~\cite{Holt10} are listed for comparison.
In fact, the realistic interaction has often been deemed as a good starting point in the construction of empirical Hamiltonians~\cite{Bro06,Honma}. It has been shown that, by correcting a few monopole (diagonal) interaction terms only, a good description on the properties of nuclei over a wide region can be obtained by calculations with realistic interactions (see, e.g., Ref.~\cite{Zuker} for a general description on the property of the monopole interaction). The origin of this monopole problem in realistic $NN$ interactions is not clear yet. It may be related to the in-medium or three-nucleon modification of the $NN$ interaction which are not taken into account in usual realistic shell-model calculations.

\begin{table}
\centering \caption{Comparisons of diagonal matrix elements $\langle ij|V|ij\rangle^{JT}$ between empirical and realistic interactions. Only the related terms are listed for simplicity. The in-medium modified interaction is taken from Ref.~\cite{Holt10}. It was calculated with the chiral  {\it NN} potential assuming $\rho=\rho_0$ where $\rho_0$ is the saturation density.} \label{table1}
\begin{ruledtabular}
\begin{tabular}{ccccccc}
&\multicolumn{2}{c}{$J=0,T=1$} &&\multicolumn{3}{c}{$J=1,T=0$} \\
\cline{2-3}\cline{5-7}
Interaction&$0p_{3/2}^2$&$0p_{1/2}^2$&&$0p_{3/2}^2$&$0p_{3/2}0p_{1/2}$&$0p_{1/2}^2$\\
\hline
CK~\cite{ck}&-3.19&-0.26&&-3.58&-6.22&-4.15\\
HWM~\cite{van88}&-3.68&-0.15&&-2.62&-6.55&-3.95\\
WBT~\cite{wbt}&-3.85&-1.22&&-4.16&-6.86&-3.45\\
WBP~\cite{wbt}&-3.91&-1.15&&-3.86&-6.94&-3.45\\
Chiral~\cite{Holt10}&-3.28&-0.61&&-1.19&-5.20&-1.67\\
Medium Chiral~\cite{Holt10}&-0.12&~1.91&&~1.29&-3.03&~0.72\\
\end{tabular}
\end{ruledtabular}
\end{table}

One may expect the in-medium modification was approximated by simple approaches like the  Brown-Rho scaling. However, it is known that the  Brown-Rho in-medium modification may lead to serious
changes of the property of the effective interaction. The drastic effects of the Brown-Rho mass scaling of different
baryons on $sd$-shell effective interactions were firstly discussed in Ref.~\cite{Hosaka}. It is supported by our systematic calculations in different model spaces with various {\it NN} potentials employing the G-matrix and other renormalization techniques~\cite{Qi08,Qi09}. Calculations of Refs.~\cite{Holt08,Holt09,Holt10} in the $p$-shell share the same problem, as seen in Table~\ref{table1}. For examples, the interaction terms $\langle 0p_{1/2}^2|V|0p_{1/2}^2\rangle^{J=0,T=1}$ and $\langle 0p_{1/2}^2|V|0p_{1/2}^2\rangle^{J=1,T=0}$ change from $-0.61$~MeV and $-1.67$~MeV to 1.91~MeV and 0.72~MeV, respectively, by applying the in-medium modification on the chiral  {\it NN} potential~\cite{Holt10}. This kind of strange interaction strengths is not supported by any existing interactions constructed for this region nor experimental data. In fact, it can be easily recognized that these matrix elements should be attractive by looking at the binding energies and one-neutron separation energies of C and N isotopes~\cite{Audi03}.

The empirical interactions of Refs.~\cite{ck,van88,wbt} are constructed in the particle-particle channel by assuming $^{4}$He as the inert core and the single-particle energies as free parameters. But in all cases it was assumed that the interactions are are same in the hole-hole channel. The energy splitting between $0p_{1/2}^{-1}$ and $0p_{3/2}^{-1}$ orbits in $^{15}$C and $^{15}$N is calculated to be $\varepsilon=\varepsilon(0p_{3/2}^{-1})-\varepsilon(0p_{1/2}^{-1})=6.3$~MeV~\cite{ck}, 7.3~MeV~\cite{van88} and 6.5~MeV~\cite{wbt}.
The interaction of Ref.~\cite{Holt10} is evaluated directly in the hole-hole channel with the energy splitting being taken as experimental datum, i.e, $\varepsilon=6.3$~MeV.

The ground state of $^{14}$C is dominated by the configuration of $|0p^{-2}_{1/2}\rangle $ due to the large spin-orbit splitting between orbits $0p_{1/2}^{-1}$ and $0p_{3/2}^{-1}$. This is supported by calculations with both empirical and realistic interactions, as seen from Table~\ref{table2}.  As a result, 
the coefficient $\kappa$ of Eq.~(1) is significantly larger than $\eta$
($\kappa$ and $\eta$ have the same sign and $\kappa=1$ in the single-particle
limit). The mixing of the two corresponding configurations is induced by the non-diagonal matrix element $\langle 0p_{3/2}^2|V|0p_{1/2}^2\rangle^{J=0,T=1}$ for which realistic and empirical interactions give a similar strength.

Similarly, one may safely expect that  $|0p^{-2}_{1/2}\rangle $ should be the dominated configuration in the ground state wave function of $^{14}$N since the other two configurations lie at much higher energies. This expectation is supported by all calculations listed in Table~\ref{table2}. 

Since the amplitude $\kappa$ is much larger than $\eta$,  the suppression of the Gamow-Teller transition
strength [Eq. (2)] should be largely due to the cancellation between $a$ and $2b$
($a$ and $b$ have different signs). That is, the term $|0p_{3/2}^{-1}0p_{1/2}^{-1}\rangle$ should most likely be the second largest component in the ground state wave function of $^{14}$N. Most of our calculations with different interactions predict that the absolute value of the amplitude of $|0p_{3/2}^{-2}\rangle$ is small. In this case we should have $a\sim-(2\kappa+\sqrt{2}\eta)b$ in reproducing the cancellation. As seen from Table~\ref{table2}, the problem of the realistic calculations of Ref.~\cite{Holt10} is that the predicted amplitude $a$ ($b$) is significantly smaller (larger) than expected.

\begin{table}
\centering \caption{Comparisons of wave functions calculated with empirical and realistic interactions. All calculations are done with the code~\cite{Qi08a} except those of Jancovici and Talmi's and of the chiral potential which are taken from Ref.~\cite{Jan54} and Ref.~\cite{Holt10}, respectively.} \label{table2}
\begin{ruledtabular}
\begin{tabular}{ccccccc}
Interaction&$\eta$&$\kappa$&&$c$&$b$&$a$\\
\hline
CK~\cite{ck}&0.38&0.92&&-0.027&-0.31&0.95\\
HWM~\cite{van88}&0.36&0.93&&-0.063&-0.27&0.96\\
WBT~\cite{wbt}&0.31&0.95&&~0.033&-0.43&0.90\\
WBP~\cite{wbt}&0.30&0.95&&0.014&-0.41&0.91\\
JT~\cite{Jan54}&0.09&0.99&&~~0.20&-0.41&0.89\\
Zamick~\cite{Zam66}&0.22&0.98&&0.014&-0.40&0.92\\
VF~\cite{Vis57}&0.25&0.97&&0.12&-0.36&0.97\\
Chiral~\cite{Holt10}&0.40&0.92&&~~0.14&-0.68&0.72\\
Medium Chiral~\cite{Holt10}&0.26&0.97&&~~0.11&-0.59&~0.80\\
\end{tabular}
\end{ruledtabular}
\end{table}

The ratio between the amplitudes $a$
and $b$ are sensitive to the strengths of diagonal interaction matrix elements $\langle 0p_{1/2}^2|V|0p_{1/2}^2\rangle^{J=1,T=0}$ and $\langle 0p_{3/2}0p_{1/2}|V|0p_{3/2}0p_{1/2}\rangle^{J=1,T=0}$ and the non-diagonal matrix element $\langle 0p_{3/2}0p_{1/2}|V|0p_{1/2}^2\rangle^{J=1,T=0}$. If we neglect the contribution from the configuration $|0p_{3/2}^2\rangle$ and restrict the calculation to dimension two, the mixing between above two components would be solely dominated by the term $(\varepsilon+\langle 0p_{3/2}0p_{1/2}|V|0p_{3/2}0p_{1/2}\rangle-\langle 0p_{1/2}^2|V|0p_{1/2}^2\rangle)/\langle 0p_{3/2}0p_{1/2}|V|0p_{1/2}^2\rangle$. Bearing this in mind, one may inspect the in-medium modification calculations of Refs.~\cite{Holt08,Holt09,Holt10}. It is interesting to see that the few involved non-diagonal matrix elements are not much affected by the Brown-Rho or three-nucleon modifications (This may be consistent with the fact that the non-diagonal parts of modern empirical and realistic interactions are similar to each other). For their calculations with the Bonn-B potential, which reproduced the cancellation under certain conditions, the relative strength of above two diagonal matrix elements are rectified at the cost of changing dramatically their absolute values~\cite{Holt09,Holt10}.

Still it may be of interest to figure out the role played by the tensor force in effective interactions in the $jj$ coupling scheme (The impossibility of inducing cancellation with effective interactions without tensor force was first shown by Inglis~\cite{Ing53}).  Although they are mixed in usual construction of empirical interactions, it is possible to separate the central, spin-orbit (vector) and tensor force components of the effective interaction through the spin-tensor decomposition procedure~\cite{Kir73}.
As examples in Table~\ref{table3} is listed the different components of the effective interactions of Refs.~\cite{van88,wbt}. The decompositions of various Cohen-Kurath interactions can be found in Ref.~\cite{Yoro80} and will not be presented here for simplicity. Similar decompositions are also done for realistic interactions derived from the state-of-the-art Bonn potential~\cite{Machleidt}.
But as pointed out in Ref.~\cite{Kir73}, there is no trivial relation between the spin-tensor decomposition of the effective interaction and different components of the underlying $NN$ potential. Especially, the tensor force of the $NN$ potential may contribute significantly to the overall effective interaction when renormalization and core-polarization effects are taken into account~\cite{Kuo65}.

\begin{table}
\centering \caption{The central, spin-orbit (SO) and tensor components of the matrix elements $\langle ij|V|kl\rangle^{JT}$ of empirical interactions~\cite{van88,wbt}.} \label{table3}
\begin{ruledtabular}
\begin{tabular}{cccccccc}
&\multicolumn{3}{c}{HWM} &&\multicolumn{3}{c}{WBT} \\
\cline{2-4}\cline{6-8}
$ijkl$&Central&SO&Tensor&&Central&SO&Tensor\\
\hline
\multicolumn{8}{c}{~~~~~~~~~~~~~~~~~~~~~$J^{\pi}=0^+,T=1$}\\
1111&-1.66&0.58&0.92&&-1.33&054&-0.42\\
3333&-4.43&0.29&0.46&&-3.96&0.33&-0.21\\
1133&-3.93&-0.41&-0.65&&-3.72&-0.42&0.29\\
\hline
\multicolumn{8}{c}{~~~~~~~~~~~~~~~~~~~~~$J^{\pi}=1^+,T=0$}\\
1111&-4.27&~~0.41&-0.098&&-4.49&1.11&-0.075\\
1113&~1.08&~-0.21&0.83&&0.43&-0.0019&1.38\\
1313&-5.85&~~0.10&-0.80&&-5.85&0.35&-1.36\\
1133&~2.72&~-0.13&-0.46&&1.55&-0.046&-0.82\\
1333&~3.72&0.065&-0.016&&2.41&0.045&-0.012\\
3333&-2.96&0.041&0.30&&-4.11&-0.58&0.53\\
\end{tabular}
\end{ruledtabular}
\end{table}

In agreement with the earlier calculation of Ref.~\cite{Yoro80}, it is seen that the central force components of realistic and empirical interactions are similar to each other. Noticeable differences are only seen in certain matrix elements of the non-central spin-orbit and tensor forces. 

In Table~\ref{table4} is listed the wave functions calculated with the tensor force component removed from the effective interactions. In this case it is seen that the ground state wave function of $^{14}$N is overwhelmingly dominated by the configuration of $|0p^{-2}_{1/2}\rangle $. This is because the non-diagonal  matrix element $\langle 0p_{3/2}0p_{1/2}|V|0p_{1/2}^2\rangle^{J=1,T=0}$, which is crucial in inducing the configuration mixing, is dominated by the contribution from the tensor force. As mentioned before, the non-diagonal matrix elements are resistant to the Brown-Rho and three-nucleon in-medium modifications. Systematic comparisons between realistic and empirical effective interaction in this work as well as other studies~\cite{Zuker,Bro06,Honma} tend to suggest that these non-diagonal matrix elements are well defined by the underlying $NN$ potential. The problem of realistic interactions in reproducing the cancellation should be related to its uncertainties in the few diagonal monopole terms.

\begin{table}
\centering \caption{Wave functions of $^{14}$C and $^{14}$N calculated with the central force and central and spin-orbit force components of effective interactions.} \label{table4}
\begin{ruledtabular}
\begin{tabular}{ccccccc}
Interaction&$\eta$&$\kappa$&&$c$&$b$&$a$\\
\hline
\multicolumn{7}{c}{~~~~~~~~~~~~~~~~~~~~~Central force only}\\
CK~\cite{ck}&0.33&0.94&&-0.15&-0.085&0.99\\
HWM~\cite{van88}&0.29&0.96&&-0.14&-0.086&0.99\\
WBT~\cite{wbt}&0.30&0.95&&-0.11&-0.033&0.99\\
\hline
\multicolumn{7}{c}{~~~~~~~~~~~~~~~~~~~~~Central plus spin-orbit}\\
CK~\cite{ck}&0.34&0.94&&-0.15&-0.085&0.99\\
HWM~\cite{van88}&0.32&0.95&&-0.15&-0.053&0.99\\
WBT~\cite{wbt}&0.33&0.94&&-0.12&-0.030&0.99\\
\end{tabular}
\end{ruledtabular}
\end{table}

Summarizing, systematic calculations with a variety of shell-model effective interactions predict consistent results on the wave functions of the ground states of $^{14}$C and $^{14}$N, both of which are dominated by the configuration of $0p_{1/2}^{-2}$. It is seen that the accidental cancellation of the $^{14}$C-dating $\beta$ decay amplitude is largely induced by the mixing effect of two configurations of the final state wave function, $|0p^{-2}_{1/2}\rangle$ and
$|0p_{3/2}^{-1}0p_{1/2}^{-1}\rangle$. The mixing between these two components are sensitive to a few diagonal matrix elements and one non-diagonal matrix element which is mainly determined by the tensor force. The failure  of realistic calculations in reproducing
the inhibition may be related to its ill description of the monopole component rather than the tensor force. 
A rigorous correction of the monopole interaction
by more comprehensive in-medium modification or other microscopic
approaches would be useful. Work in this direction is underway.

\section*{ACKNOWLEDGMENTS}
This work has been supported by the Swedish Research Council (VR) under contract No. 623-2009-7340. I would like to thank Mr. X.B. Wang (Beijing) for his help.

\section*{Appendix}

In many cases the wave functions of $^{14}$C and $^{14}$N were calculated in the $LS$ coupling scheme~\cite{Jan54}.
The transformation between wave functions in $LS$ and $jj$ coupling schemes is known in analytic forms in terms of $6j$ and $9j$ symbols. To facilitate the comparison between wave functions in different coupling schemes available on the market, the explicit expressions for the transformation are listed below as
\begin{equation}
\left(
  \begin{array}{c}
    |^1S_0\rangle \\
    |^3P_0\rangle \\
  \end{array}
\right) = \frac{1}{\sqrt{3}}\left(
            \begin{array}{cc}
              1 & \sqrt{2} \\
              \sqrt{2} & -1 \\
            \end{array}
          \right)\left(
                            \begin{array}{c}
                              |0p^{-2}_{1/2}\rangle \\
                              |0p_{3/2}^{-2} \rangle\\
                            \end{array}
                          \right),
\end{equation}
and
\begin{equation}
\left(
  \begin{array}{c}
 |^3S_1\rangle \\
|^1P_1\rangle \\
|^3D_1\rangle \\
  \end{array}
\right) = \frac{1}{\sqrt{3}} \left(
            \begin{array}{ccc}
              -1 & -4 & \sqrt{10} \\
              \sqrt{6} & \sqrt{6} & \sqrt{15} \\
              \sqrt{20} & -\sqrt{5} & -\sqrt{2} \\
            \end{array}
          \right)\left(
                            \begin{array}{c}
                            |0p^{-2}_{1/2}\rangle \\
                              |0p_{3/2}^{-1}0p_{1/2}^{-1}\rangle \\
                              |0p_{3/2}^{-2}\rangle \\
                            \end{array}
                          \right).
\end{equation}

\end{document}